\documentclass[showpacs,preprintnumbers,amssymb]{revtex4}

\usepackage{pstricks}%
\usepackage{graphicx}
\usepackage{dcolumn}
\usepackage{bm}
\usepackage{dsfont}
\usepackage{amsmath}
\usepackage{pifont}
\usepackage{psfrag}
\usepackage{hyperref}
\usepackage[latin2]{inputenc}
\usepackage{bbm}

\renewcommand\Re{\text{Re\,}}
\renewcommand\Im{\text{Im\,}}

\newcommand{\BE}{\begin{equation}}
\newcommand{\EE}{\end{equation}}

\newcommand{\skipc}[2]{}

\newcommand{\fig}[1]{Fig.~\ref{#1}}
\newcommand{\eq}[1]{Eq.~(\ref{#1})}

\newcommand{\I}{\ensuremath{{\mkern1mu\mathrm{i}\mkern1mu}}}
\newcommand{\E}{\ensuremath{{\mkern1mu\mathrm{e}\mkern1mu}}}

\newcommand{\sep}{\ensuremath{{\mkern1mu\mathrm{sep}\mkern1mu}}}

\newcommand{\qed}{\nobreak \ifvmode \relax \else
      \ifdim\lastskip<1.5em \hskip-\lastskip
      \hskip1.5em plus0em minus0.5em \fi \nobreak
     $\square$\fi}

\newcommand{\be}{\begin{equation}}
\newcommand{\ee}{\end{equation}}
\newcommand{\bea}{\begin{eqnarray}}
\newcommand{\eea}{\end{eqnarray}}

\newcommand{\ket}[1]{\ensuremath{|#1\rangle}}
\newcommand{\bra}[1]{\ensuremath{\langle#1|}}

\usepackage{color}

\begin{document}

\title{Characterizing the width of entanglement}

\author{Sabine W\"olk}
\affiliation{Naturwissenschaftlich-Technische Fakult\"at, Universit\"at Siegen, Walter-Flex-Str.~3, 57068 Siegen, Germany}
\author{Otfried G\"uhne}
\affiliation{Naturwissenschaftlich-Technische Fakult\"at, Universit\"at Siegen, Walter-Flex-Str.~3, 57068 Siegen, Germany}

\date{\today}

\begin{abstract}
The size of controllable quantum systems has grown in recent times. Therefore, the spatial degree of  freedom becomes more and more important in experimental quantum systems. However, the investigation of entanglement in many-body systems mainly concentrated on the number of entangled particles and ignored the spatial degree of freedom, so far. As a consequence, a general concept together with experimentally realizable criteria have been missing to describe the spatial distribution of entanglement. We close this gap by introducing the concept of entanglement width as measure of the spatial distribution of entanglement in many-body systems. We develop criteria to detect the width of entanglement  based solely on  global observables. As a result, our entanglement criteria can be applied easily to many-body systems since single-particle addressing is not necessary.
\end{abstract}

\pacs{3.67.Mn, 3.56.-w, 5.50.+q}

\maketitle

\section{ Introduction}
To experimentally implement quantum technologies \cite{Haeffner2008,Raimond2001}, such as information processing, simulation \cite{Georgescu2014} or metrology \cite{Toth2014},  characterizing and understanding multipartite entanglement \cite{Horodecki2009} is important. By characterizing multipartite entanglement one is able to  understand experimental setups, identify experimental limitations, and investigate possible noise sources in a better way. 
The  size of controllable  quantum systems has grown in recent times and large arrays of atoms \cite{Birkl2014} or clouds of macroscopic singlet states \cite{Mitchell2014} have been produced. Also ideas of coupling several ion traps to build a quantum computer  exist \cite{Wunderlich2015}. However, for large quantum systems the spatial degree of freedom  becomes more and more important since for large systems external fields cannot be approximated by constant fields anymore. Furthermore, entanglement can be easily protected again constant fields but is very vulnerable to spatially varying fields. As a result, the spatial distribution of entanglement compared to the spatial distribution of external fields is decisive for the temporal evolution of spatially extended quantum systems. 

The spatial distribution of entanglement plays also an important role in the investigation of quantum phase transitions \cite{Osterloh2002, Osborne2002,Richerme2014b,Gu2004,Biswas2014,Hofmann2014} and makes a distinction  between different given ground states of the generalized Heisenberg spin-chain possible \cite{Eckert2008, Chiara2011}.

Entanglement of multipartite system can be characterized with different 
quantities, such as the entanglement depth or $k$-producibility 
\cite{Sorensen2001, Guehne2005}, which is defined as the minimum number 
of entangled particles necessary to create a given state. 
In systems without any spatial ordering, entanglement depth is a powerful 
variable to characterize the entanglement properties of this system. 
However, in  systems with spatial order such as spin chains, or in 
the presence of gradient fields, the dynamics of a system may also 
depend on whether entanglement  exists only between neighboring 
particles or between distant particles.  

The investigations done so far concentrated on entanglement depth 
or $k$-producibility \cite{Sorensen2001,Guehne2005,Guehne2006} or 
required  addressability of single subsystems \cite{Osterloh2002, Osborne2002, Richerme2014b,Gu2004,Biswas2014}.  Our criteria, developed in this paper, are based solely on global observables. Therefore, they open the possibility to study 
correlation propagation and other physical characteristics of 
many-body systems without the necessity of addressing single subsystems.

The paper is organized as follows: First, we introduce the concept of entanglement width to characterize the spatial distribution of entanglement. Then, we give an example of how the width of entanglement influences the time evolution of a quantum system before we present methods  to characterize the width of entanglement with the help of global observables. We conclude by demonstrating how quantum phase transitions manifest themselves in the width of entanglement.


\begin{figure}[t]
\includegraphics[width=0.4\textwidth]{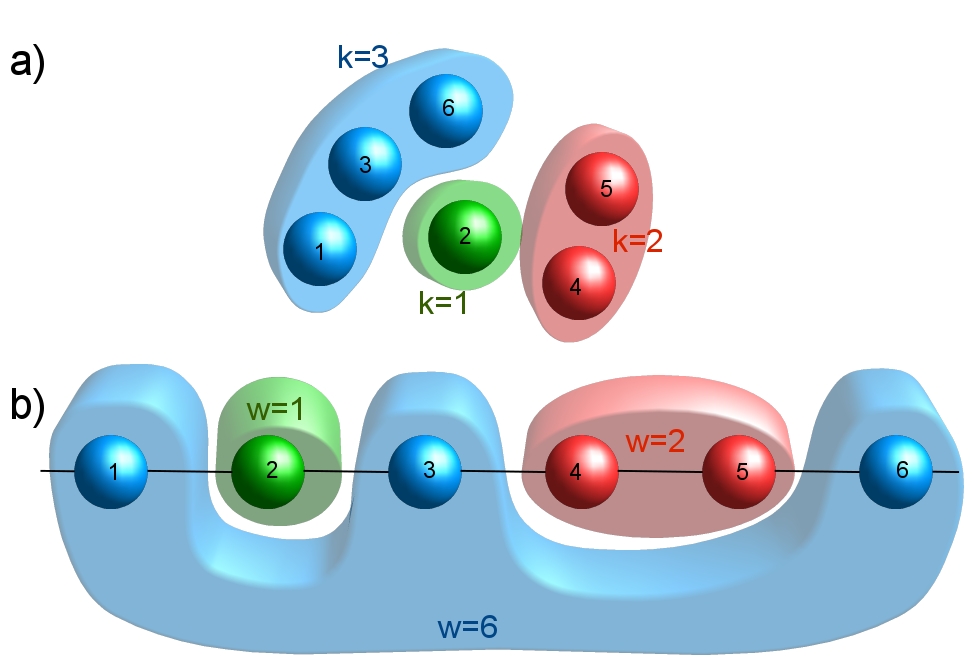}\\
 \caption{Comparison of entanglement depth (a) and entanglement 
 width (b) of the state $\ket{\Psi}=\ket{\psi_{1,3,6}}\otimes \ket{\psi_{4,5}}\otimes\ket{\psi_{2}}$. Whereas the entanglement 
 depth disregards any spatial ordering, the definition of entanglement 
 width requires the particles to be spatially ordered, e.g. in a 
 spin chain. The entanglement depth of the state  $\ket{\Psi}$ in (a)
 is given by $k=3$ (since maximally three particles are entangled). This
 is a lower bound on the entanglement width in (b), which  equals $w=6$
 (since entanglement occurs over a distance of six particles in the chain).}
 \label{fig:ent_width}
\end{figure}

\section{The concept of entanglement width}
 
The width of entanglement $w$ of a pure state $\ket{\Psi}=\bigotimes_j \ket{\psi_j}$ is defined as the maximal distance $w$ of two entangled particles within the states $\ket{\psi_j}$ (see \fig{fig:ent_width}). A completely separable state exhibits an entanglement width of $w=1$. The entanglement width of a mixed state is defined by the minimum with $w$ over all decomposition $\varrho=\sum_j p_j \ket{\psi_j}\bra{\psi_j}$, that is 
 \BE
 w(\varrho)=\underset{\textrm{decompositions}}{\textrm{min}}\Big[\underset{j}{\textrm{max }}\{w(\psi_j)\}\Big].
 \EE
By definition, the entanglement depth is a lower bound of the entanglement width. However, the entanglement width does not make any statement about the entanglement depth. For example, the width of entanglement $w=6$ in \fig{fig:ent_width} stays the same, no matter if all particles are entangled with each other or only the two outer ones (1 \& 6) whereas the entanglement depth changes from $k=6$ to $k=2$.
Furthermore, states with equal entanglement depth but different entanglement width can lead to dramatic different effects as we demonstrate in the following example.

\begin{figure}[t!]
   \begin{center}
   \includegraphics[width=0.95\textwidth]{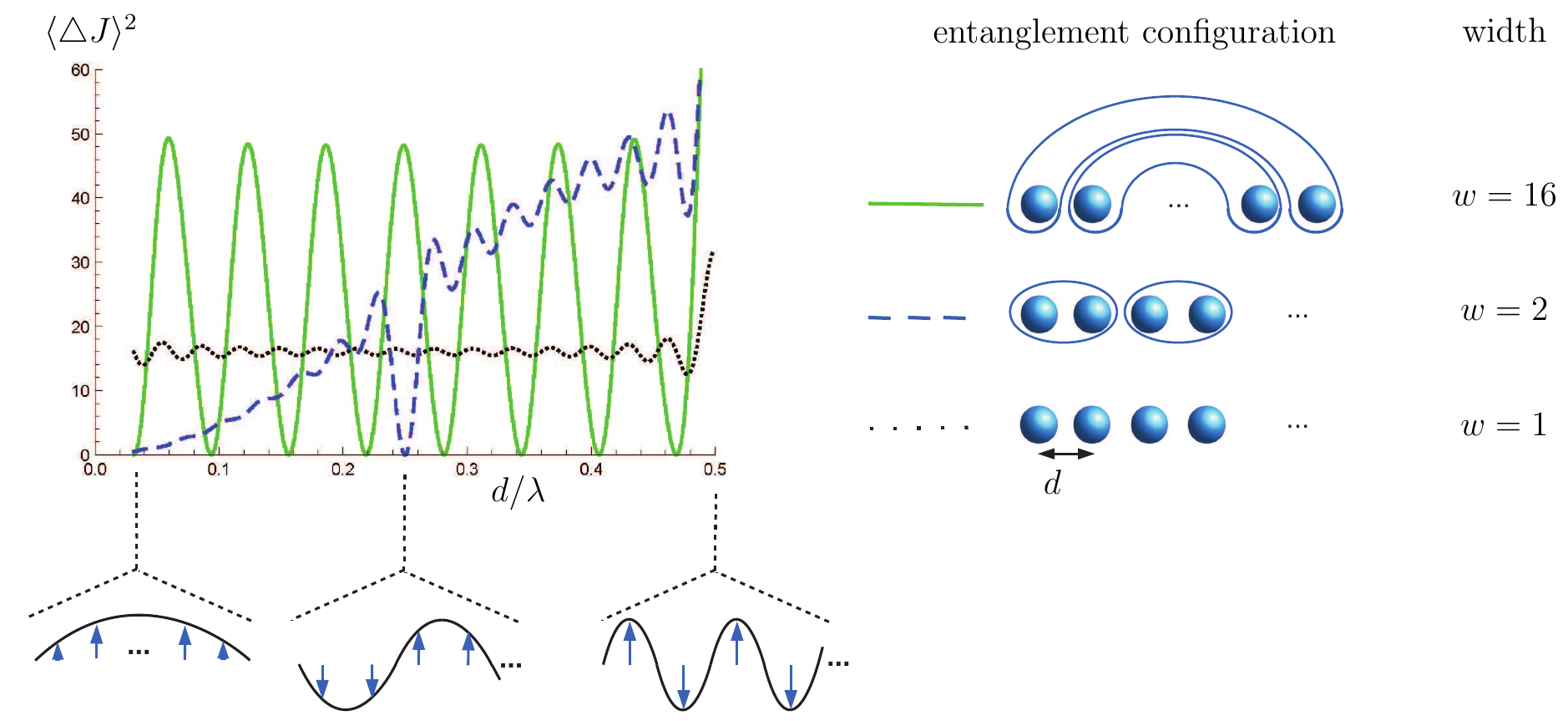}
   \end{center}
\caption{  Variance of the observable $\vec J$ for different parameter $\lambda$ and entanglement configurations for $N=16$ particles. Pairs of encircled particles form together the state $\ket{\psi^-}$. The product state (non-encircled particles, $w=1$) is chosen in such a way, that it minimizes the variance. Although, both entangled states possess the same entanglement depth $k=2$, they exhibit quite different behavior due to their different spatial distribution of entanglement.\label{fig:overview}}
\end{figure}

Motivated by Ref. \cite{Eckert2008} consider a chain of particles $j$ at the positions $x_j=x_0+j\cdot d$, with equal spacing $d$  between two particles. Furthermore, consider the observable
\BE
\vec{J}=\sum\limits_{j=1}^N \sin\left(2\pi \frac{x_j}{\lambda}\right) \vec{\sigma}_j\label{eq:def_J}
\EE
with $\vec{\sigma}_j=\left(\sigma^{(x)}_j,\sigma^{(y)}_j,\sigma^{(z)}_j\right)^\dagger$ denoting the Pauli matrices acting on particle $j$ and $\lambda$ being a parameter determining the observable. Such an observable can be created e.g. by a time evolution
\BE
U(\frac{d}{\lambda})=\exp\big[2\pi\I \sum\limits_j \frac{x_j}{\lambda}\sigma_j^{(y)}\big]\label{eq:time_evo}
\EE
 of spins in a gradient field which rotates the spins depending on their position. Here, $\lambda$ is given by the gradient of the field. 
Another approach to create an observable similar to $\vec{J}$ is given by a standing light wave coupled to cold atoms in a lattice.  In this way, Eckert et al. were able to distinguish between 4 given ground states with different spatial entanglement configuration which are important for condensed matter and high energy physics \cite{Eckert2008}.
Whereas Eckert et al. needed the previous knowledge that their state is given by exactly one out of four states, the following method to investigate the width of entanglement does not need this strict restriction. 

$\vec{J}$ is given by the global spin operator, if $d/\lambda$ is an integer and e.g. $x_0=\lambda/4$. Its variance $(\Delta \vec{J})^2=(\Delta J_x)^2+(\Delta J_y)^2+(\Delta J_z)^2$, with $(\Delta J_x)^2=  \langle J_x^2\rangle - \langle J_x\rangle^2$, is minimized by the global singlet state
\BE
\ket{\Psi^-}=\bigotimes_{(j,k)}\ket{\psi^-}_{j,k}
\EE
were particles are combined to pairs forming together the singlet state $\ket{\psi^-}=(\ket{01}-\ket{10})/\sqrt{2}$.
However, for $d/\lambda \notin \mathbb{N}$ the variance depends on the spatial distribution of the entangled pairs $(j,k)$ as can be seen in \fig{fig:overview} for $N=16$ and $x_0=-d/2$. For example for $d/\lambda = 1/(2N)$ the ``hugging'' configuration with $w=N$,  where particle $j$ is entangled with particle $N+1-j$ (solid green line), reaches the minimal variance of $(\Delta \vec{J})^2_\text{hug}=0$. However, if all particle with odd number $j$ are entangled with their right neighbor $k=j+1$ (dashed blue line),  the width of entanglement is given by $w=2$, which leads to the variance 
\BE
(\Delta \vec{J})^2_\text{rn}=\frac{3}{2} N (1-\cos(\frac{\pi}{N}))\approx \frac{3\pi^2}{4N}.\label{eq:rn}
\EE
As a consequence, the variance of $\vec{J}$ depends crucially on the spatial distribution of entanglement. 

Also the quantum Fisher information (QFI), which is an important measure for quantum metrology and entanglement  \cite{Hyllus2012}, is strongly influenced by the spatial distribution of entanglement. For example, the time evolution given in \eq{eq:time_evo} leads to a QFI given by $F_\text{hug}= N^4$ for the hugging configuration, whereas it is given by $F_\text{rn}=4N^2$ for the right neighbor configuration. As a consequence, the two here considered states exhibit different behavior although their are equal in their entanglement depth.



\section{ First criterion for entanglement width}
\begin{figure}
   \begin{center}
   \vspace*{0.6em}
   \includegraphics[width=0.5\textwidth]{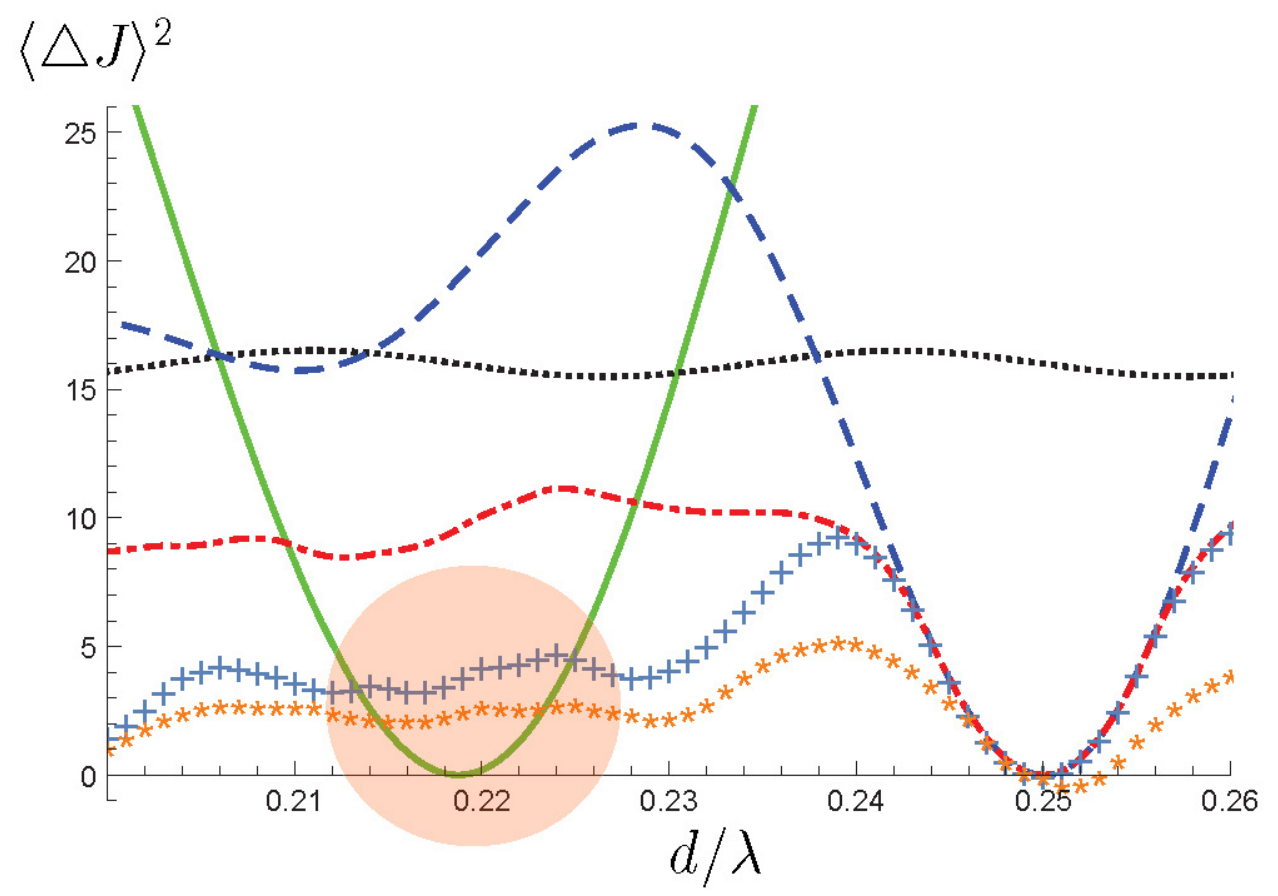}
   \vspace*{-1em}
   \end{center}
\caption{  Variance of the observable $\vec J$ for different parameter $\lambda$ and entanglement configurations and $N=16$ particles. Lines in (green, solid), (blue, dashed) and (black, dotted) correspond to the states given in \fig{fig:overview}. The limits for certain entanglement configurations are given by (i) (red, dashed-dotted) :lower limit for the entanglement configuration $(1,2),\dots (15,16)$, (ii) (blue, $+$): lower bound for $w=2$, (iii) (orange, $\ast$ ): lower bound for $w=4$. The state in the hugging configuration with width $w=16$ (green, solid) violates in the encircled area the lower bounds for entanglement width $w=2$ and $w=4$. Therefore its long range entanglement is detected by the presented method. \label{fig:lowerbounds}}
\end{figure}

In this section we demonstrate that the observable $\Delta \vec{J}$, defined in  \eq{eq:def_J}, is able to distinguish between short-range and long-range entanglement. We will estimate the minimal variance $(\Delta \vec{J})^2$ for states with nearest-neighbor entanglement and demonstrate that states possessing long range entanglement are able to violate these bounds. Since $(\Delta \vec{J})^2$ is a concave function, it reaches its minimum for pure states. Furthermore, since nearest-neighbor entanglement implies the entanglement of maximal two particles, we find for pure states
$
(\Delta \vec{J})^2=\sum_{(j,k)}(\Delta \vec{J})^2_{(j,k)}
$
with the two-particle variance
\BE
(\Delta \vec{J})^2_{(j,k)}= 3(a_j^2+a_k^2)-\left[\left(a_j\langle \vec{\sigma}_j\rangle +a_k\langle\vec{\sigma}_k\rangle\right)^2-2a_ja_k\langle \vec{\sigma}_j\vec{\sigma}_k\rangle\right]\label{def:2pvariance}
\EE
where $a_j=\sin\left(2\pi x_j/\lambda\right)$. The minimum of this two-particle variance is given by
\BE
\underset{\ket{\psi}}{\text{min}}(\Delta \vec{J})^2_{(j,k)}=\left\lbrace \begin{array}{cl}
 a_j^2\left(2+2\varepsilon^2-\frac{4\varepsilon^2}{(1-\varepsilon)^2}\right)& -1\leq \varepsilon\leq \varepsilon_0\\
3a_j^2(1-\varepsilon)^2& \varepsilon_0\leq \varepsilon \leq 1
\end{array}
\right.
\EE
where we assumed with out loss of generality  $|a_j|>|a_k|$ and defined $\varepsilon=a_k/a_j$ and $\varepsilon_0=2-\sqrt{3}\approx 0.27$ (for a proof see Appendix A). The remaining task is to optimize over all possible combination of entangled pairs. This is a classical optimization task and for many parameter $\lambda$ easy to estimate. For example for $\lambda=1/(2N)$ and $N=4k+2$ with $k\in \mathbb{N}$ the optimal pairing is given by $(1,2)..,(N-1,N)$ and we find $\varepsilon>2-\sqrt{3}$ for all pairs of $a_j,a_k$. Therefore, the lower bound is exactly given by \eq{eq:rn}. A simple lower bound for the optimal pairing is given by
\BE
\underset{\lbrace(j,k)\rbrace}{\text{min}}(\Delta \vec{J})^2\geq \sum\limits_{j=1}^N \frac{1}{2} \underset{k}{\text{min}}(\Delta \vec{J})^2_{j,k}.
\EE
In this way, we can find easily lower bounds not only for nearest-neighbor entanglement but also for general maximal widths $w$. For example, in \fig{fig:lowerbounds} we show for $N=16$ the variance $(\Delta \vec{J})^2_\text{hug}$ which can beat the limits for $w=2$ and $w=4$ for $\lambda/d\approx 7/(2N) $.

\section{  Second criterion for entanglement width}
Another method to distinguish between short range and long range entanglement is to use several observables with different correlations. For example the Hamiltonian 
\BE
\hat H_1=\sum\limits_{j=1}^N \vec{\sigma}_{j} \vec{\sigma}_{j+1}.\label{eq:H1}
\EE 
of a spin chain in the Heisenberg model contains only nearest neighbor correlations.  On the other hand, the total collective angular moment 
$\vec{J}_c$ \eq{eq:def_J} with $d/\lambda \in \mathbb{N}$
 includes correlations between all spins with equal weight. Both observables can be used to detect multipartite entanglement \cite{Hyllus2012,Toth2009,Guehne2005}. However, they can only together distinguish between nearest-neighbor and long range entanglement.

The  difference between the energy and the total spin may be a good indicator for entanglement beyond nearest-neighbors, which implies an entanglement width of $w\geq 3$. Indeed, by defining the correlation function 
\begin{eqnarray}
\chi(N)&\equiv& \langle J_c^2-2H\rangle-3N \nonumber\\
&=&\sum\limits_j\sum\limits_{k=2}^{N-2}\vec{\sigma}_{j} \vec{\sigma}_{j+k} \label{eq:chi}
\end{eqnarray}
one can define a correlation measurement, which includes only non-nearest neighbor correlations. 

For quantum states with only nearest-neighbor entanglement  the correlations appearing in  $\chi(N)$ can all be treated classically. The minimum of $\chi(N)$ can then be derived with the help of circulant matrices \cite{gray2006} (see 
Appendix B). As a consequence, all states with entanglement width $w\leq 2$ satisfy the inequality
\BE
\chi(N)\geq -N\frac{\sin(3\pi/N)}{\sin(\pi/N)}.\label{theo:h1}
\EE
For large $N$ we find
\BE
 -N\frac{\sin(3\pi/N)}{\sin(\pi/N)}\xrightarrow[N\gg 1]{}-3N+\frac{4\pi^2}{N}.
\EE
A quantum state, which has obviously entanglement beyond nearest-neighbor entanglement is the state
\BE
\ket{\Psi}_{nnn}=\bigotimes\limits_{k=0}^{N/4-1}\ket{\psi^-}_{k+1,k+3}\ket{\psi^-}_{k+2,k+4},
\EE
where always two non-nearest-neighbor spins form together the Bell state $\ket{\psi^-}=(\ket{01}-\ket{10})/\sqrt{2}$. 
As a consequence, we find $\chi _{nnn}(N)=-3N$ which violates for all finite $N$ the bound $-N \sin(3\pi/N)/\sin(\pi/N)$. The violation decreases with $1/N$. However, note that a maximal violation proportional to $1/N$ appears also in other entanglement criteria, see e.g. Ref.~\cite{Sorensen2001b}. The reason may be, that our criterion is highly symmetric and the minimal overlap between maximal entangled symmetric states and product states decreases faster than $1/N$ \cite{Aulbach2010}, which is a consequence of the de Finetti theorem.


As an application for this criterion, we consider spin chains with nearest-neighbor 
and next-nearest-neighbor coupling described by the Hamiltonian
\BE
H_2=\sum_{j=1}^N \vec{\sigma}_j\vec{\sigma}_{j+1} + \alpha \sum_{j=1}^N \vec{\sigma}_j\vec{\sigma}_{j+2}.\label{eq:H2}
\EE
This model is often called the $J_1$-$J_2$-model. It is used to understand 
phenomena in magnetic materials such as Tomonaga-Luttinger-liquids states 
and spin-Peierls states \cite{Haldane1982, Somma2001}. Here, the interactions between nearest and next-nearest neighbor compete with each other,  leading 
to frustrated spins.

To estimate the quality of our criterion, we compare it with a criterion for entanglement depth.
For separable state the minimal energy $\langle H_2 \rangle_\sep$ can be computed with the help of the eigenvalues of the circulant correlation matrix. With this method, we determine for separable states
$ \langle H_2\rangle_\sep \geq N h_\textrm{circulant}$
with 
\BE
h_\textrm{circulant}=\underset{m}{\textrm{min}}\left[\cos(2\pi\frac{m}{N})+\alpha\cos(2\pi\frac{2m}{N})\right]\label{eq:hcirculant}.
\EE
To estimate the minimal achievable energy $H_2$ for states with entanglement depth $k\leq 2$ we use the methods from Refs.~\cite{Guehne2005, Guehne2006} (see Appendix~C). As a consequence, for  states with entanglement depth $k\leq 2$ we find the limit
$\langle H_2\rangle_\textrm{2-prod}\geq -N h_{2\textrm{prod}}\label{eq:2prod}$
with
\BE
h_{2\textrm{prod}}=1+\alpha+\frac{1}{2+4\alpha}.\label{eq:h2prod}
\EE

As an example we investigate the entanglement of the ground state  of  a spin chain with $N=8$ spins and an interaction Hamilton given by $H_2$ defined in \eq{eq:H2} for different values of $\alpha$ \footnote{The ground state is evaluated numerically with the help of the QUBIT4MATLAB packages \cite{Toth2008}.}.

\begin{figure}[t]
\includegraphics[width=0.65\textwidth]{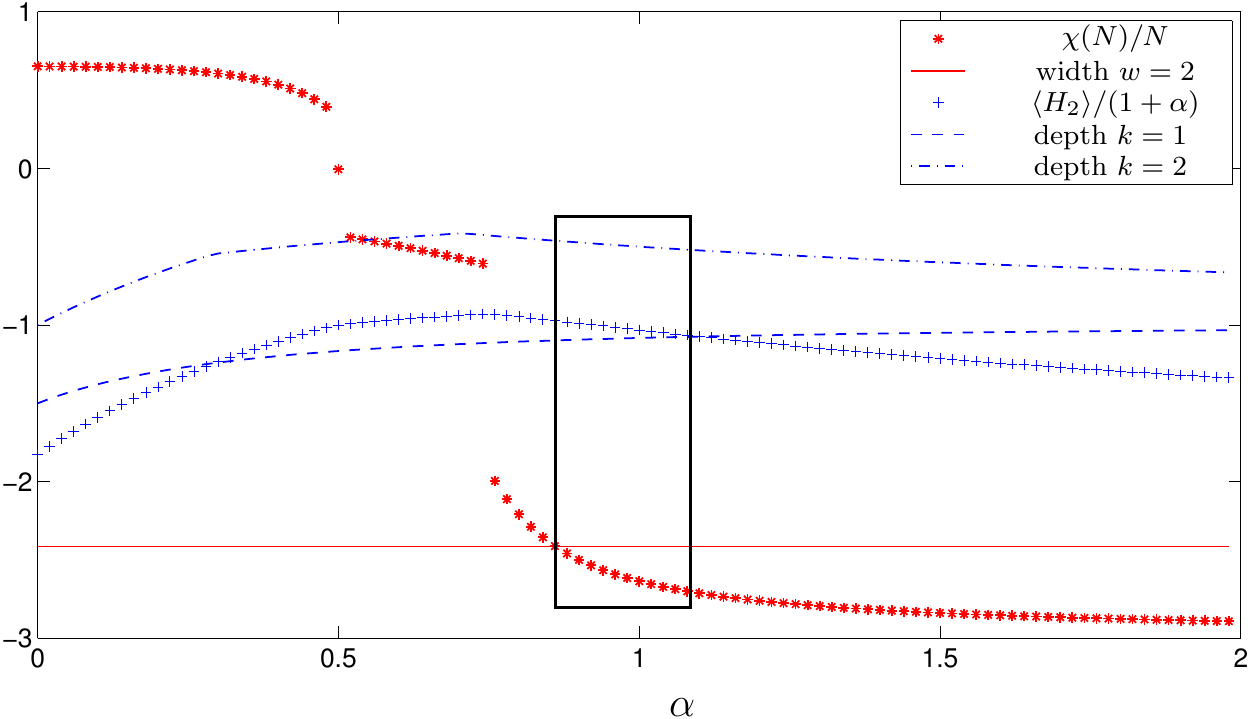}
\caption{Comparison of entanglement depth and entanglement width  for different $\alpha$ and $N=8$ spins: the energy $H_2$ of the ground state (blue ``$+$'') indicates entanglement if it lies below $h_\textrm{circulant}$ (blue dashed-dotted line) and multipartite entanglement if it lies below $h_\textrm{2prod}$  (blue dashed line). The expectation value of $\chi_2$ (red ``$\star$'')  indicates entanglement beyond nearest neighbor if it lies below $\chi_\textrm{class}$ (red line). As a consequence, our results detect entanglement of non-neighboring particle before we detect multipartite entanglement (black square). \label{fig:ent_H2}}
\end{figure}

In \fig{fig:ent_H2} we compare the expectation value of the energy $H_2$ (blue crosses) with the minimal achievable energy for states with entanglement depth $k\leq 1$ (blue dashed dotted line) and $k\leq 2$ (blue dashed line) as well as the correlation function $\chi$ (red $\ast$) and the limit for states with entanglement width $w\leq 2$ (red line).
As can be seen in \fig{fig:ent_H2}  the ground state exhibits entanglement depth of $k\geq 3$ for $\alpha\lesssim 0.3$ and $\alpha\gtrsim 1.1$. On the other hand, we detect  entanglement width of $w\geq 3$ for $0.86  \lesssim \alpha$ using $\chi$. As a consequence, there exist a regime (black square) where we detect already an entanglement width of $w>2$ with the help of the correlation function $\chi$, but no multipartite entanglement with existing methods. Furthermore, the correlation function $\chi$ shows two sudden jumps at $\alpha\approx 0.5 $ and  $\alpha\approx 0.7 $ which are indicators that for $N \rightarrow \infty$  phase transitions may occur at these points \cite{Gu2004}. Contrary to that, we find no hint for phase transitions in the expectation value $\langle H_2 \rangle$.

The behavior found with our  entanglement criteria fits  well to existing results in  the literature. As shown by Majumdar and Ghosh \cite{Majumdar1969,Majumdar1969b}, the ground state of $H_2$ is given by
\BE
\ket{\psi_{MG}}=\bigotimes\limits_{k} \ket{\psi^-}_{2k,2k\pm1}
\EE
for $\alpha=1/2$. Therefore, the ground state at this point shows bipartite entanglement between nearest neighbors as also indicated by our results. Furthermore, the relation between entanglement and phase transitions for the Hamiltonian $H_2$ has been investigated with the help of the concurrence \cite{Gu2004} and  generalized geometric measures \cite{Biswas2014}. Similar to our results, Gu et~al. found with the help of the concurrence  that the entanglement changes from nearest neighbor to next-nearest neighbor entanglement around $\alpha\approx 0.75 $ for $N=8$ spins \cite{Gu2004}. Biswas et~al. showed that phase transitions which occur only for $N\rightarrow \infty$ manifest themselves in the  generalized geometric measure for entanglement also for finite $N$ around   $\alpha\approx 0.7 $ \cite{Biswas2014}.   Our entanglement criterion is easily accessible experimentally in contrast to the methods used by Biswas et~al. and Gu et~al., which need the exact knowledge of the state and addressability of 
single particles.


\section{ Conclusion}
In summary, we have introduced  the concept of entanglement width and demonstrated that states with equal entanglement depth but different entanglement width behave differently. 
We developed criteria based solely on global observables which are able to distinguish between different values of entanglement width. With the help of these entanglement criteria we investigated the ground state of a spin chain described by $H_2$ and showed that we are able to detect long range entanglement before we detected multipartite entanglement. 

Furthermore, phase transitions as predicted by other theories for spin-chains described by $H_2$  manifest  themselves in our concept of entanglement width whereas they were not visible in our measure for entanglement depth.   Although, we just started to investigate the width of entanglement it turned out to be an important indicator to investigate the behavior of quantum systems in the presence of space dependent interactions. In this way, we could used it as indicator for many-body phenomena such as phase transitions. It would be very interesting to further investigate this 
possibility. 

In addition, detecting the width of entanglement with global observables is not limited to the two methods demonstrated in this paper. Another example is e.g. given by measuring the variance of the total spin with and without a linear dependency on the position and a consecutive comparison. Here, the variance of the total spin without external field gives information about the entanglement, whereas the position dependent observable measures the spatial distribution of correlations, classical as well as quantum (for more details see Appendix D). We are certain that many more criteria detecting the width of entanglement will be developed in the future.

We thank G. T\'oth, R. Sewell, J. Kong and M. Mitchell for fruitful discussions. This work has been supported by the EU (Marie Curie CIG 293993/ENFOQI),  the FQXi Fund (Silicon Valley Community Foundation), the DFG and the ERC (Consolidator Grant 683107/TempoQ).

\begin{appendix}

\section{The minimal two-particle variance}
The two particle variance $(\Delta \vec{J})^2_{(j,k)}=1+\varepsilon^2-g(\varepsilon,\ket{\psi})$ with
\BE
g(\varepsilon,\ket{\psi})=\left(\langle \vec{\sigma}_j\rangle +\varepsilon\langle\vec{\sigma}_k\rangle\right)^2-2\varepsilon\langle \vec{\sigma}_j\vec{\sigma}_k\rangle
\EE
is minimal if $g$ riches its maximum. To maximize g we parametrizes the state $\ket{\psi}$ by
\BE
\ket{\psi}=a\ket{\psi^-}+b\ket{\psi^+}+c\ket{\phi^-}+d\ket{\phi^+}.
\EE
with the usual definition of the Bell states $\ket{\psi^-}$, $\ket{\psi^+}$, $\ket{\phi^-}$ and $\ket{\phi^+}$.
As a consequence, the spin vector of the single particle is give by
\BE
\langle \vec{\sigma}\rangle_{j/k}=2\left(\begin{array}{c}
\Re(bd^\ast)\mp\Re(ac^\ast)\\ \Im(b^\ast c)\pm\Im(ad^\ast)\\\Re(c^\ast d)\pm\Re(ab^\ast)
\end{array}\right).
\EE
Since the two vectors always lie in a plane, we assume $\langle \sigma_y \rangle=0$ for both particles. This implies phase relations between the parameters $a-d$ which implies that the maximum of $g$ can be reached by choosing all parameters to be real. As a consequence we get
\BE
g(\varepsilon,\ket{\psi})=4(b^2+c^2)d^2(1+\varepsilon)^2+ 4(b^2+c^2)a^2(1-\varepsilon)^2
+8\varepsilon a^2-2\varepsilon
\EE
which does not depend on $b^2$ or $c^2$ itself but only on their sum. Therefore, we choose w.o.l.g  $b=0$ and arrive at
\BE
g(\varepsilon,\ket{\psi})=8a^2\varepsilon(1-2c^2)-4c^4(1+\varepsilon)^2-2\varepsilon
\EE
where we also made use of the normalization relation $a^2+b^2+c^2+d^2=1$ of the state. The function $g$ is positive and monotonically  increasing in $a^2$ for  $\varepsilon>0$ and $c^2<1/2$ and negative for $\varepsilon>0$ and $c^2>1/2$. Therefore, the maximum of $g$ for $\varepsilon>0$ is reached by choosing the maximal possible value of $a^2$ given by $a^2=1-c^2$. Now, we are able to maximize $g$ over the only left parameter $c^2$. The maximum is reached for 
\BE
c^2=\frac{1}{2}-\frac{\varepsilon}{(1-\varepsilon)^2}.\label{eq:c}
\EE
However, this is only possible for $\varepsilon < 2-\sqrt{3}$ since $c^2\geq 0$ . If $\varepsilon > 2-\sqrt{3}$, we have to choose $c^2=0$ and $a^2=1$ to maximize $g$. As a consequence, for $\varepsilon\geq 2-\sqrt{3}\approx 0.27$ the variance $\triangle \vec{J}$ is minimized by the singlet state $\ket{\psi^-}.$

In a similar way, the maximum of $g$ for $\varepsilon<0$ can be estimated and we find that choosing again $c$ as defined in \eq{eq:c} is optimal. However,  for $-1\leq \varepsilon \leq 0$ we find $1/2 \leq c^2 \leq 3/4$  which makes now further distinction of cases necessary.


\section{ Lower bound of $\chi(N)$ for $w=2$}

For $w=2$ the correlation function $\chi(N)$ defined in Eq.(5) becomes classical. Therefore, it
can be written as  $\chi^\textrm{cl}(N)=\vec{X}^T C\vec{X}$, with $\vec{X}=(\vec{x}_1,\dots,\vec{x}_N)^T$, $\vec{x}_j = \langle \vec{\sigma}\rangle $ and the circulant correlation matrix C.
As a result, the minimum of $\chi_N^\textrm{cl}$ for only nearest-neighbor entangled states is bounded from below by the minimal eigenvalue of $C$. The correlation matrix $C$ is a circulant matrix $[16]$, and
the eigenvalues of $C$ are given by
\BE
\lambda_m=\sum\limits_n c_n \E^{2\pi \I \frac{mn}{N}}=\left\{\begin{array}{cl} N-3&\textrm{ for }m=0\\-\frac{\sin(3\pi\frac{m}{N})}{\sin(\pi\frac{m}{N})} & \textrm{ for } m=1,\dots,N-1\end{array}\right.
\EE
with $c_n$ denoting the entries of the circulant matrix $C$.
 As a consequence, the minimal eigenvalue of $C$ is given by $\textrm{min}(\lambda_m)=-\sin(3\pi/N)/\sin(\pi/N)$.  

To estimate the minimum of $\chi^\textrm{cl}(N)$ we have to consider also the length of the eigenvector $\vec{X}_m$. Since $|\vec{x}_j|\leq 1$ the maximal length of the vector $\vec{X}$ is given by $N$. Indeed, by choosing $\vec{X}=(\vec{x}_1,\dots,\vec{x}_N)^T$ with e.g.
\BE
\vec{x}_j=\left(\begin{array}{c}\cos 2\pi \frac{j}{N}\\\sin 2\pi \frac{j}{N}\\ 0\end{array}\right)
\EE
we have found an eigenvector of $C$ with the minimal eigenvalue and the length $N$.

We note two interesting facts: (i) the eigenvectors of circulant matrices like $C$ are independent of the coefficient $c_k$, only the eigenvalues depend on $c_k$. (ii) Although the vectors $\vec{x}_j$ live in a three-dimensional space, the vectors $\vec{x}_j$ forming the eigenvector are only two-dimensional. Since the eigenvectors are independent of the coefficient $c_k$ the scheme presented here can be generalized to all correlation functions
\BE
\chi=\sum\limits_j\sum\limits_k c_{j,k}\vec{x}_j\vec{x}_k
\EE
which are circulant, that is $c_{j,k}=c_{|j-k|}$ and the minimum of $\chi$ is always given by $N\cdot \textrm{min}(\lambda)$. As a consequence, the scheme presented here can be used to find the minimal energy of arbitrary circulant Hamiltonians.


\section{ Minimal energy $H_2$ for $k=2$}
 To derive the minimal Energy $\langle H_2\rangle $ achievable by states with entanglement depth $k \leq 2$ we divide the spin chain into blocks, where the ions within a block may be entangled but ions belonging to different blocks must be separable analogue to Refs.~$[11,12]$. Using the methods of Ref.~$[12]$, the minimal achievable energy is bounded by
\BE
\langle H_2\rangle \geq -\sum_j C_j,
\EE
where $C_j$ describes the optimization over each single block. For single-particle blocks $C_j$ is given by
\BE
C_j^\textrm{single}=\underset{\ket{\psi}}{\textrm{max}}\left( \vec{x}_k^2+\alpha \vec{x}_k^2\right)=1+\alpha.
\EE
For two-particle blocks we obtain with the help of Lemma A1 of Ref.~$[12]$
\begin{eqnarray}
C_j^\textrm{double}&=&\underset{\ket{\psi}}{\textrm{max}}\left[ - \vec{x}_k\vec{x}_l+\frac{1}{2}( \vec{x}_k^2+\vec{x}_l^2)+\alpha( \vec{x}_k^2+\vec{x}_l^2)\right]\nonumber\\
&=&2(1+\alpha)+\frac{1}{1+2\alpha}.
\end{eqnarray}
As a consequence, for  states with entanglement depth $k\leq 2$ we derive the limit
$\langle H_2\rangle_\textrm{2-prod}\geq -N h_{2\textrm{prod}}$
with
\BE
h_{2\textrm{prod}}=1+\alpha+\frac{1}{2+4\alpha}.
\EE


\begin{figure}[t]
   \begin{center}
   \vspace*{0.6em}
   \includegraphics[width=0.65\textwidth]{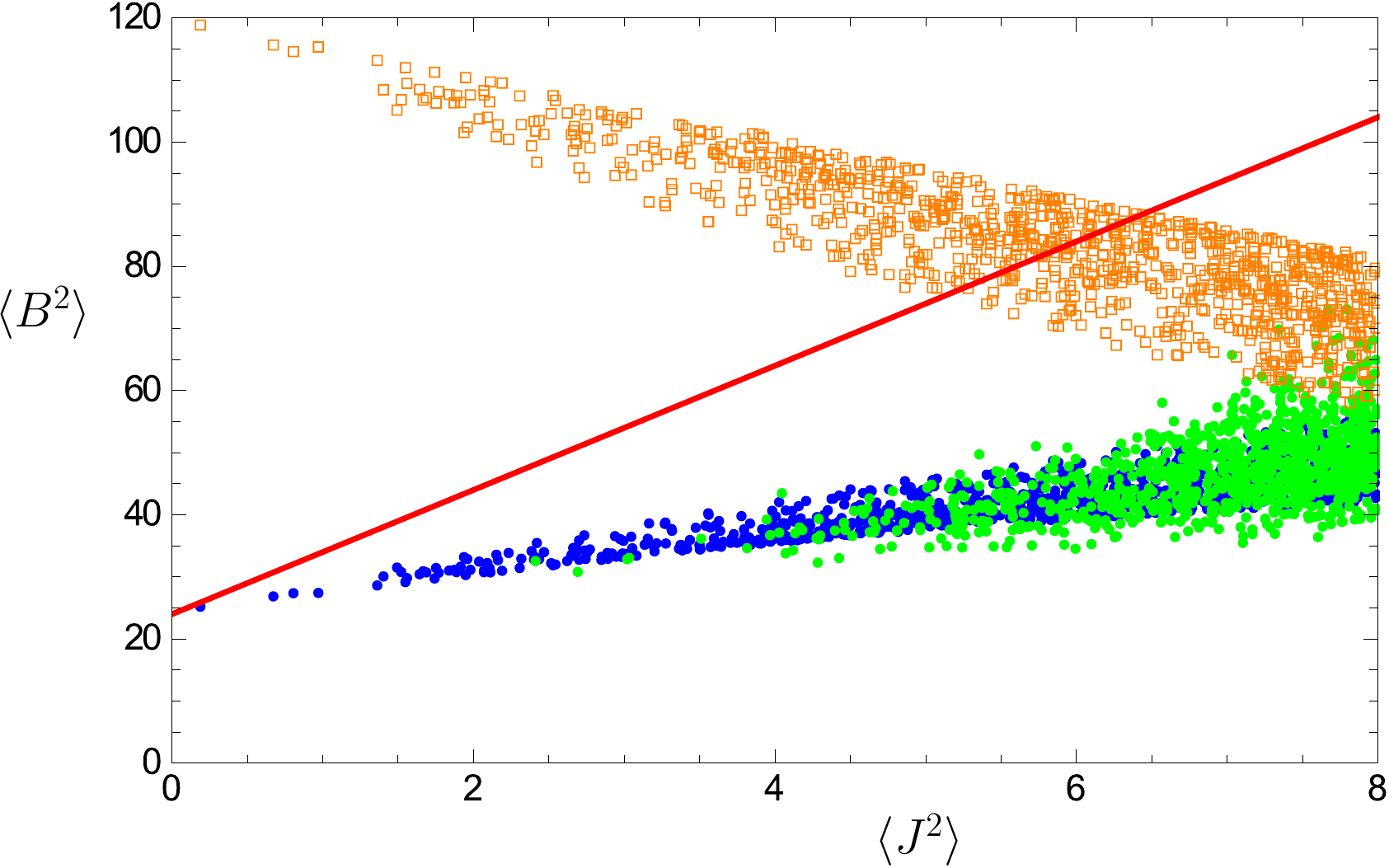}
   \vspace*{-1em}
   \end{center}
\caption{Expectation values of $J^2$ and $B^2$ for states of the form: $\ket{\psi}_{1,2}\ket{\psi}_{3,4}$ (\textcolor{blue}{$\bullet$}), $\ket{\psi}_{1,2}\ket{\phi}_{3,4}$ (\textcolor{green}{$\bullet$}), and
$\ket{\psi}_{1,4}\ket{\psi}_{2,3}$ (\textcolor{orange}{$\square$}). 
States which are separable under the partition $1,2|3,4$ lie below the limit: $\langle B^2\rangle = 24+10 \langle J^2\rangle$ (red line)  \label{fig:limit_sdp_N4}}
\end{figure}

\section{ Detecting the width of entanglement with the help of a gradient}

Another way to detect the entanglement width is to change the weight of different interactions through an external space-dependent field, e.g. by  measuring the collective spin in the presence of a magnetic gradient. In the following, we assume for simplicity an even number of spins and that the magnetic field is zero in the middle of the chain. Therefore, the interaction of the magnetic field with particle $k$ can be described by
\BE
B^{(\ell)}_{k}=(2k-N-1) \sigma^{(\ell)}_{k},
\EE
with $\ell=x,y,z$.
Similar to the angular momentum we define $B^{(\ell)}=\sum_k B^{(\ell)}_{k}$ and $B^2= (B^x)^2+(B^y)^2+(B^z)^2$.  Whereas small values of  $\langle J^2 \rangle$ are indicators for entanglement, large values of $\langle B^2 \rangle$ are indicators for long range correlations, both classical and quantum. As a consequence, all four-qubit states which are separable under the partition $1,2 | 3,4$ obey the inequality
\BE
\langle B^2 \rangle \leq 24+10\langle J^2\rangle. \label{theo:mag_grad}
\EE
as depicted in \fig{fig:limit_sdp_N4}.
This bound is tight, since it is reached at the point $(\langle J^2\rangle=0,\langle B^2\rangle=24 )$ by the state $\ket{\psi^-}_{1,2}\ket{\psi^-}_{3,4}$ and at the point $(\langle J^2\rangle=8,\langle B^2\rangle=104 )$  by the state \mbox{$\ket{\uparrow\uparrow}_{1,2}\ket{\downarrow\downarrow}_{3,4}$}. Mixtures of these two states lie exactly on the line.

\eq{theo:mag_grad} can be proven with the help of semi-definite programming. Here, we utilize the fact that all states which are separable under a given partition possess  a positive partial transposition (PPT) under this partition \cite{Peres1996, Horodecki1996}.  Therefore, we prove  \eq{theo:mag_grad} by searching the minimum of $10\langle J^2 \rangle - \langle B^2 \rangle $ for four-qubit states  $\varrho$ under the condition that the partial transpose of $\varrho$ under the partition $1,2 | 3,4$ is positive semidefinite.

However, also states which are separable under the partition $1|2,3|4$ possess only nearest-neighbor entanglement, too. For these states, an inequality similar to \eq{theo:mag_grad} can be formulated. Here, we search the minimum $-a$ of $m\langle J^2 \rangle - \langle B^2 \rangle $ for four-qubit states  $\varrho$ which are PPT under the partitions $1|2,3,4$ as well as $1,2,3|4$ for different $m$. In this way we get a family of inequalities
$
\langle B^2 \rangle \leq a+m \cdot \langle J^2\rangle.
$
In contrast to \eq{theo:mag_grad}, there exist no optimal inequality: either we optimize the y-intercept $a$ or the slope $m$.  Furthermore, we made a strong relaxation by going from separability under the partition $1|2,3|4$ to PPT under the partitions $1|2,3,4$ and $1,2,3|4$. Therefore, the gained inequalities are not tight. Nevertheless, we find joined inequalities for states which are PPT under the partition $1,2|3,4$ or simultaneous PPT under $1|2,3,4$ and $1,2,3|4$ for example $\langle B^2 \rangle \leq 24+16.1 \cdot \langle J^2\rangle$ (optimal y-intercept) or $\langle B^2 \rangle \leq 50.2+10 \cdot \langle J^2\rangle$ (same slope as in \eq{theo:mag_grad}). 
All states which violate at least one of these inequalities  possess entanglement beyond nearest neighbors.

Similar bounds can be found for $N=6$. For example we find the bound $\langle B^2 \rangle \leq 36+48\langle J^2\rangle$ for states which are simultaneous PPT under the partition $1,2|3,4,5,6$ and $1,2,3,4|5,6$. Unfortunately, such optimization problems are very complicated for large systems and can therefore not be solve with semi-definite programming for large $N$. However, for $\langle J^2\rangle=0$ and $\langle J^2\rangle=2N$ we are able to analytically calculate the upper bound of $\langle B^2 \rangle$, leading to a conjecture about the general behavior:

The only states with $\langle J^2 \rangle=0$ and only next-neighbor entanglement are given by 
\mbox{$\ket{\Psi^-}=\bigotimes\ket{\psi^-}_{2j-1,2j}$}
which leads to $\langle B^2\rangle_{-}= 6N$. The state 
\mbox{$ \ket{\Psi^\textrm{cl}}=\ket{\uparrow,\dots,\uparrow}_{1,\dots, N/2}\ket{\downarrow,\dots, \downarrow}_{N/2+1,\dots,N} $}
possess the highest possible value of $\langle B^2 \rangle$ for states with only  entanglement  within the subgroups $\lbrace 1,\dots, N/2\rbrace$ and  $\lbrace N/2+1,\dots, N\rbrace$. This state is characterized by the expectation values $\langle J^2\rangle_\textrm{cl}=2N$ and
\mbox{$\langle B^2 \rangle_\textrm{cl}=(N^4)/4+2(N-1)N(N+1)/3$}. Therefore, all states with expectation values $\langle B^2 \rangle$ exceeding this bound must posses entanglement between these two subgroups. If we allow entanglement only between nearest neighbors, then compared to $\langle B^2 \rangle_\textrm{cl}$ only the correlation between the two middle particles within the chain are allowed to change from two to six. As a consequence, all states with $\langle B^2 \rangle > \langle B^2 \rangle_\textrm{cl}+4$ posses entanglement beyond nearest neighbors.  

Furthermore, the behavior of $\langle B^2 \rangle$ for $N=4$ leads us to the conjecture, that the largest possible value of $\langle B^2\rangle$ for a given $\langle J^2\rangle$ with only  entanglement within within the subgroups $\lbrace 1,\dots, N/2\rbrace$ and  $\lbrace N/2+1,\dots, N\rbrace$  is reach by the state
\mbox{$\varrho=p \ket{\Psi^-}\bra{\Psi^-}+ (1-p)\ket{\Psi^\textrm{cl}}\bra{\Psi^\textrm{cl}}$}.
As a result, we obtain the conjecture that all states with entanglement only within the subgroups $\lbrace 1,\dots, N/2\rbrace$ and  $\lbrace N/2+1,\dots, N\rbrace$  satisfy the inequality
 \BE
 \langle B^2 \rangle \leq \langle B^2\rangle_- + \frac{\langle B\rangle_\textrm {cl}-\langle B^2\rangle_-}{2N}\,\langle J^2\rangle.
 \EE
For $N=4$ the conjecture coincides with the inequality given in \eq{theo:mag_grad}. Furthermore, first tests with random states confirm this bound. 

\end{appendix}



\end{document}